# Geometric Interpretation of Standard Model Hierarchies and Zero Neutrino Mass Scale

E. M. Lipmanov
40 Wallingford Road # 272, Brighton MA 02135 USA

**Abstract**

Numerous confirmations may indicate that, besides the problem of Higgs sector many external parameters, the Standard Model is a complete low energy particle interaction theory not crying for new physics. But there are serious problems with understanding flavor in SM phenomenology. Firstly, equal number '3' of SM particle generations coincide with the dimension number of outer euclidean 3-space and secondly, the factual SM particle mass and mixing hierarchies at leading approximation appear solutions of the Metric equation for unit vector direction angles in euclidean 3-space geometry. The power of the Metric equation in hierarchy phenomenology is visualized by explanation of the outstanding in the SM large empirical Tau-lepton, Top-quark, Bottom-quark masses and small neutrino Reactor mixing angle. New in the SM weak interactions particle color degree of freedom (3 quark colors and 1 lepton color) appears appropriate for representing quark and neutrino mixing angle unity at basic level and substantial difference at the observational level. Based on suggested by data idea of zero neutrino mss scale, absolute neutrino masses $m_1 \cong 0$, $m_2 \cong 0.009$ eV, $m_3 \cong 0.05$ eV are analyzed. They are related to the neutrino mixing angles and are of the same hierarchy type that the known Dirac mass patterns of charged leptons and quarks.

## 1. Introduction

After the Higgs boson discovery, LHC results support further the Standard Model of known elementary particles. Higgs sector is the least advanced one in the SM. There are two ways of developments – theoretical and semi-empirical. The main content of the latter is looking for physically meaningful regularities of the many empirical parameter ratios and hierarchies of masses and mixing angles in particle flavor groups of the SM Higgs sector.

The main goals of the present research are 1) unification of all known SM lepton and quark mass and mixing hierarchies at leading approximation of the appropriate Higgs parameters by universal bimaximal form hierarchy patterns and 2) geometric interpretation of these universal hierarchies. These goals are achieved in the framework of semi-empirical phenomenology. Two empirical coincidences – a) three SM particle generations coincide with three dimensions of outer euclidean 3-space, and b) introduced earlier empirical universal Pythagorean equation for neutrino mixing angles coincide with Metric equation of the outer euclidean 3-space geometry – are count as important indications of a connection between SM flavor hierarchies and space geometry. In accordance, it appears that all SM flavor hierarchies at leading (benchmark) approximations are described by solutions without external parameters of the Metric-Pythagorean equation for double hierarchy angles.

In Sec.2, euclidean 3-space Metric equation for unit vector direction angles is outlined. In Sec.3, Metric-Pythagorean equation for neutrino mixing angles is discussed. In Sec.4, quark-neutrino mixing analogy is considered. In Sec.5, the empirical quark-lepton complementarity QLC rule is outlined. In Sec.6, charged lepton and quark united universal Dirac mass hierarchies are revealed. In Sec. 7, zero neutrino mass scale is analyzed. Sec.8 contains Conclusions.

## 2. Metric equation. Three basic solutions for direction angles

Consider a constant unit vector **A** in euclidean 3-space. Its direction angles $\varphi_x$, $\varphi_y$ and $\varphi_z$ obey the euclidean symmetry Metric equation

$$\cos^2 \varphi_x + \cos^2 \varphi_y + \cos^2 \varphi_z = 1 \qquad (1)$$

in an orthogonal coordinate system with axes **X, Y, Z**. By relation (1), only two direction angles are independent free parameters. Equation (1) is protected by euclidean geometric symmetry in the 3-dimensional space.

There are only three basic 'bimaximal' visualized solutions of the metric equation (1) without external parameters:

1) Vector **A** parallel to the **Z**-axis

$$\cos^2 \varphi_k = (0, 0, 1),\ k = x, y, z,\ \varphi_x = \varphi_y = \pi/2,\ \varphi_z = 0. \qquad (2)$$

2) Two independent solutions with vector **A** orthogonal to the **Z**-axis

$$\cos^2 \varphi_k = (0, 1, 0), \varphi_x = \varphi_z = \pi/2, \varphi_y = 0, \quad (3)$$

$$\cos^2 \varphi_k = (1, 0, 0), \varphi_y = \varphi_z = \pi/2, \varphi_x = 0. \quad (4)$$

With vector **A** orthogonal to the **Z**-axis, there is geometric symmetry of the two solutions**.**

All other (infinite number) solutions of the Metric equation (1) are superpositions of these three basic equations (2), (3), (4) with free external parameters[1] $\varphi_x, \varphi_y, \varphi_z$.

An interesting derivative solution without external parameters (degenerate solution) of the Metric equation (1) is a symmetric superposition of all three bimaximal solutions (2), (3), (4):

$$\cos^2\varphi_k = (1/3, 1/3, 1/3), \varphi_k \cong 54.7^o . \quad (5)$$

It should be noted that there are two different geometric symmetry types of solutions without free external parameters of Eq. (1): the system of three solutions (2)-(4) and solution (5) are geometrically symmetric, but each of the three solutions (2)-(4) if separated violates geometric symmetry.

The interesting physical fact revealed in the present research is that these two types of solutions are pertinent for describing the mixing angle patterns of the two known different types of lepton and quark mixing in the SM. The primary leading approximation neutrino mixing is described (Sec. 3) by mixing angle geometric symmetry violating bimaximal solution (2), while the leading approximation mixing angles of individual color quarks is described by geometrically symmetric solution (5) in order to get the indicated by data zero mixing solution at leading approximation after summing up all three quark colors (Sec. 4).

## 3. Neutrino mixing angles

1) Pythagorean equation for neutrino angles

Inspired by the experimental indications of not equal zero neutrino theta_13 mixing angle [2], Pythagorean equation for neutrino double mixing angles ($\theta_{12}$, $\theta_{23}$, $\theta_{13}$ are solar, atmospheric and reactor neutrino mixing angles respectively) was proposed in 2011 arXiv publication ref [1],

$$\cos^2 2\theta_{12} + \cos^2 2\theta_{23} + \cos^2 2\theta_{13} = 1, \quad (6)$$

---

[1] The metric equation (1) is linear with respect to the quantities $\cos^2\varphi_k$.

as semi-empirical geometric type explanation of the new phenomenon[2]**.** Indeed, since it was already well known that the solar angle is not maximal, the inference $\theta_{13} \neq 0$ immediately follows from Eq (6). Only two of the three neutrino mixing angles in Eq (6) are independed. If one angle is small ($<< \pi/4$), both other two angles must be large; if two angles are large, the third one must be small. Zero, two or three small mixing angles are forbidden by geometry. It is exactly what is observed experimentally for neutrino mixing pattern in contrast to the quark one.

But more accurate experimental data [3]-[7] for the three neutrino mixing angles indicated that Eq (6) is not likely an exact one at least at 2-3 $\sigma$ accuracy of the three mixing angles. So, a definitely compatible with data statement − Pythagorean Eq (6) is a physical relation for data neutrino mixing angles at leading approximation represented by an intrinsic geometric solution of that equation without free external parameters conformable to solution (2).

Thus in fair agreement with known experimental data [7], the unique benchmark approximation solution of Pythagorean Eq (6) for neutrino mixing angles is bimaximal solution

$$\theta_{12} = \theta_{23} = \pi/4, \ \theta_{13} = 0. \qquad (7)$$

That solution of Eq. (6) coincides with solution (2) of the metric Eq. (1) with $\varphi_x = 2\theta_{12}$, $\varphi_y = 2\theta_{23}$, $\varphi_z = 2\theta_{13}$.

As inference from the experimental data, the true benchmark approximation for neutrino mixing angles (7) appears a spontaneous violation[3] of the geometrically symmetric solution system (2)-(4) of Metric-Pythagorean equations (1), (6) outlined explicitly in Sec. 2.

Unitary mixing matrix with angles (7) is the bimaximal matrix,

$$\begin{pmatrix} 1/\sqrt{2} & 1/\sqrt{2} & 0 \\ -1/2 & 1/2 & 1/\sqrt{2} \\ 1/2 & -1/2 & 1/\sqrt{2} \end{pmatrix}_{\nu} \qquad (8)$$

---

[2] The coefficients '2' in Eq (6) are needed to keep the term "bimaximal" in agreement with regular neutrino mixing terminology.

[3] The term 'spontaneous symmetry violation' has here its primary definition of symmetry violation by an 'external' force. The characteristic feature of this definition is that it contains the potential of symmetry restoration at some appropriate physical conditions.

This widely discussed in the literature [9] neutrino mixing matrix approximately describes empirical neutrino mixing data.

2) 'Inertia' interpretation of neutrino mixing angles

Consider free inertial motion of a flavor neutrino with constant momentum vector **P**. It is a special inertial motion because the flavor identity of these particles at free motion is changing with time – oscillation phenomena, conditioned by the mixing of neutrino mass eigenstates in the weak interactions that produce the neutrinos.

Consider the constant neutrino momentum vector **P** as a particular choice of vector **A** in Sec. 2. Then the three angles of solution (7) at leading approximation have dual physical meaning – 1) neutrino mixing angles in the weak interactions (7), on the one hand, and 2) direction angles (2) of the momentum vector **P** in the outer euclidean 3-space, on the other hand. In an arbitrary orthogonal coordinate system with axes **X, Y, Z** the direction of momentum vector **P** in an inertial frame of reference is represented by

$$\mathbf{P} = (\cos\varphi_x, \cos\varphi_y, \cos\varphi_z)P. \qquad (9)$$

Geometrically singled out choice of the coordinate system is by axis **Z** coinciding with neutrino momentum vector **P**. Then relation (9) is simplified

$$\mathbf{P}/P = (0, 0, 1),\ \ \varphi_x = \varphi_y = \pi/2,\ \ \varphi_z = 0. \qquad (10)$$

It is a solution of the Metric equation (1) with direction angles $\varphi_k$ of the momentum vector **P** related to neutrino bimaximal angles (7) $\varphi_x = 2\theta_{12}$, $\varphi_y = 2\theta_{23}$, $\varphi_z = 2\theta_{13}$.

3) Bimaximal neutrino mixing as benchmark for realistic one

It should be outlined that the experimental data on neutrino mixing angles e.g. [7] indicate that the realistic neutrino mixing angles are deviated from the bimaximal benchmark pattern (7), but not much deviated.

As illustration, those deviations are approximated here by one small ε-parameter [1] related to the fine structure constant ($\varepsilon \cong \sqrt{\alpha}$)

$$\varepsilon \cong \exp(-5/2) \cong 0.082085. \qquad (11)$$

Three simple, conformable suggested by data ε-parameterizations of the realistic neutrino mixing angles are given by

$$\cos^2(2\theta^{\nu}_{12}) = f(-2\varepsilon),\ \cos^2(2\theta^{\nu}_{23}) = f(\varepsilon^2),\ \sin^2(2\theta^{\nu}_{13}) = f(\varepsilon), \qquad (12)$$

where f(x) is a universal function,

$$f(x) = |x \exp(x)|. \qquad (13)$$

The mixing angle values (12), (13) coincide with the benchmark solution (7) at $\varepsilon = 0$.

From (12), the magnitudes of neutrino mixing angles are given by

$$\theta^{\nu}_{12} \cong 34^{\circ},\ \theta^{\nu}_{23} \cong 42.6^{\circ},\ \theta^{\nu}_{13} \cong 8.7^{\circ}. \qquad (14)$$

By comparison with recent experimental data global analysis [7],

$$\theta^{\nu}_{12} = (33.7 \pm 1.1)^{\circ},\ \theta^{\nu}_{23} = (40.7 \pm 1.7)^{\circ},\ \theta^{\nu}_{13} = (8.8 \pm 0.4)^{\circ}, \qquad (15)$$

the predicted angles (14) are in good agreement with experimental data (15) for all three neutrino mixing angles.

## 4. Quark-neutrino mixing analogy

1) Analogy between individual color quark mixing and neutrino mixing

With 3 quark colors and 1 neutrino color, complete quark-neutrino mixing analogy is represented by quark-neutrino 4-color symmetric[4] system of Metric-Pythagorean equations,

$$\cos^2 2\theta^{k}_{12} + \cos^2 2\theta^{k}_{23} + \cos^2 2\theta^{k}_{13} = 1,\ k = 0, 1, 2, 3. \qquad (16)$$

One neutrino color is at $k = 0$ and 3 quark colors are at $k = c = 1, 2, 3$.

For 3 colors, the quark color-symmetric[5] Metric-Pythagorean equation is

$$\cos^2 2\theta^{c}_{12} + \cos^2 2\theta^{c}_{23} + \cos^2 2\theta^{c}_{13} = 1,\ c = 1, 2, 3. \qquad (17)$$

The two different types of elementary particle mixing in the SM – quarks and neutrinos types – are represented here by the two different solutions (2) and (5) of Eq (1). In contrast to neutrino mixing, color-quark mixing is described not by bimaximal solution (7), but by the symmetric solution (5), namely by the degenerate solution without external parameters,

$$\cos^2 2\theta^{c}_{ij} = 1/3,\ \theta^{c}_{ij} \cong 27.4^{\circ},\ ij = 12, 23, 13,\ c = 1, 2, 3, \qquad (18)$$

the mixing angle values (18) are independent of color 'c' and angle-indexes 'ij'.

---

[4] Lepton as fourth color was first considered by Pati and Salam in a symmetry-grounded research [8].

[5] Note, the concept of particle 'color' and 3-color quark symmetry is considered here on a phenomenological ground without direct relation to QCD.

But the individual degenerate color quark mixing angles $\theta^c_{ij}$ in (18) are not observable at least at low energies. All experimental measurements are made with hadrons that are 'white' symmetric sums of color quarks. In contrast to neutrinos, there are no large quark mixing angles in the data. Zero angles at leading benchmark approximation are the data indicated choice for quarks at low energies.

Accordingly, realistic zero quark mixing angles denoted as "$\theta^q_{ij}$" at benchmark approximation are defined as 'color-white' sums over 3 quark colors from Eq (18),

$$\cos^2 2\theta^q_{ij} = \sum^c \cos^2 2\theta^c_{ij} = \sum^c 1/3 = 1\,. \qquad (19)$$

And so, it directly follows that the sum over colors in (19) leads to zero quark benchmark mixing and only in case of three colors.

The solution Eq (19) for quark mixing angles is a realistic benchmark zero mixing one,

$$\cos^2 2\theta^q_{ij} = 1,\ \theta^q_{ij} = 0,\ ij = 12, 23, 13. \qquad (20)$$

The condition of color-white-quark zero mixing (20) at leading benchmark approximation is possible only if the number of quark colors '3' is equal to the number '3' of quark flavor mass copies.

Summarizing, universal degenerate color mixing angles (18) for individual color quarks are related to hidden mixing-angle symmetry that accompanies quark color symmetry. They are not observable at least at low energy weak interactions. Observable 'white' mixing angles as summed-up effects of three color-quark mixing in (19), (20) are the approximate experimental mixing angles at low energies. They are equal zero at benchmark approximation in agreement with hadron particle data. Actually, the transition from the color quantities $\cos^2 2\theta^c$ to white ones $\cos^2 2\theta^q$ in (19) means that the three color-quark particles with equal not zero mixing angles (18) are replaced by one effective white hadron particle with zero mixing angles (20) as indicated by known data.

2) Zero quark mixing as benchmark for realistic quark mixing angles

We represent the quark deviations from benchmark zero mixing by three conformable parametrizations obtained from the respective neutrino ones (12) by evident replacements

cos() → sin() for the largest two angles only[6] (compare the benchmark quark (20) and neutrino (7) angles), and change sign of electric charge $\varepsilon \to (-\varepsilon)$:

$$\sin^2(2\theta^q_{12}) = f(2\varepsilon), \ \sin^2(2\theta^q_{23}) = f(\varepsilon^2), \ \sin^2(2\theta^q_{13}) = f(\varepsilon^4), \qquad (21)$$

Eq. (21) coincides with benchmark zero solution (20) at $\varepsilon = 0$ as needed.

At small value $\varepsilon \neq 0$ the predicted by (21) quark mixing angles are a little deviated from zero benchmark (20). Realistic values of quark mixing angles follow from Eq. (21)

$$\theta^q_{12} \cong 13.05°, \ \theta^q_{23} \cong 2.36°, \ \theta^q_{13} \cong 0.19°. \qquad (22)$$

By comparison with PDG data [10],

$$\theta^q_{12} = (13.02 \pm 0.04)°, \ \theta^q_{23} = (2.35 \pm 0.05)°, \ \theta^q_{13} = (0.20 \pm 0.01)°, \qquad (23)$$

one notices a surprisingly good agreement with predictions (22) for all three quark mixing angles especially for the Cabibbo angle $\theta^q_{12}$.

## 5. The origin of quark-lepton QLC rule

Defined in the previous Sections two different benchmark approximations for neutrino and quark mixing angles may be finely expressed as solutions of two different phenomenological equations with conformable left sides:

1) Phenomenological Metric-Pythagorean equation (6) for bimaximal neutrino mixing

$$\cos^2 2\theta^\nu_{12} + \cos^2 2\theta^\nu_{23} + \cos^2 2\theta^\nu_{13} = 1, \qquad (24)$$

2) Phenomenological equation (not geometric one) for zero quark mixing angles

$$\cos^2 2\theta^q_{12} + \cos^2 2\theta^q_{23} + \cos^2 2\theta^q_{13} = 3. \qquad (25)$$

The right side integers 1 and 3 in (24) and (25) have direct phenomenological meaning of particle colors, one color for neutrinos and three colors for quarks. Here particle color degree of freedom is essential – it unites neutrino and quark mixing at the basic level in Eq (16) and distinguishes them by the numbers of color at the phenomenological observational level in (24) and (25).

The system of phenomenological neutrino and quark mixing angle equations (24) and (25) with leading benchmark solutions – bimaximal one for neutrinos from (24) and zero one for quarks from (25) − explains naturally the known empirical indication on quark-lepton complementarity QLC-rule [11] $\theta^{\nu}_{12} + \theta^{q}_{12} \cong \pi/4$, $\theta^{\nu}_{23} + \theta^{q}_{23} \cong \pi/4$.

---

[6] The expression for the smallest mixing angle $\theta^q_{13}$ is independently chosen by condition of conformity, with the function $f(\varepsilon^4)$ further specified by data.

Indeed, at the leading benchmark approximation, $\theta^{\nu}_{12} = \theta^{\nu}_{23} = \pi/4$; $\theta^{q}_{12} = \theta^{q}_{23} = 0$, this rule is an exact one. Thus at realistic observational level it is an approximated one, e. g. (14) and (22).

The main new conclusion is that the physical origin of approximate quark-lepton complementarity QLC-rule is conditioned by the substantial difference of leptons and quarks by the numbers of color.

And more, there is an answer to the usual question of why the QLC-rule is applied only to the two largest mixing angles? The answer is essentially related to the Metric-Pythagorean equation (24) that governs neutrino mixing angles at least at leading approximation. It tells that the third angle is not free and is necessarily small if the first two neutrino angles are large, while all three quark mixing angles are definitely small.

## 6. Charged lepton and quark mass hierarchies

The power of the Metric-Pythagorean equation extends beyond particle mixing angle hierarchies to the field of charged lepton and up- and down-quark mass hierarchies. It is related to the known from experimental data mass hierarchy feature that all SM Dirac particle masses at leading approximations are nearly biminimal and the Tau-lepton mass and Top- and Bottom-quark masses are relatively very large in each of the triads (e, μ, τ),.(u, c, t) and (d, s, b).

### **1)** Charged lepton mass hierarchies

We approximate the PDG values [10] of the CL masses −

$m_e = 0.510998929 \pm 0.000000011$ MeV, $m_\mu = 105.6583715 \pm 0.0000035$ MeV, $m_\tau = 1776.82 \pm 0.16$ MeV −

by biminimal pattern

$$m_e = m_\mu = 0,\ m_\tau \neq 0. \qquad (26)$$

The CL mass hierarchy (26) should be represented by "mass-hierarchy angles" according to the definition

$$\cos^2 2\theta_e = m_e /(m_e + m_\mu + m_\tau) = 0,\ \cos^2 2\theta_\mu = m_\mu /(m_e + m_\mu + m_\tau) = 0,$$

$$\cos^2 2\theta_\tau = m_\tau /(m_e + m_\mu + m_\tau) = 1, \qquad (27)$$

$$\theta_e = 45^o, \ \theta_\mu = 45^o, \ \theta_\tau = 0, \qquad (28)$$

in exact analogy with neutrino mixing angle bimaximal benchmark pattern (7).

It should be noted that the definition (27) for mass hierarchy angles is certainly favorable because a) it is e-μ-τ symmetric, b) it involves only relevant masses and no external parameters and c) those angles exactly obey the Metric-Pythagorean equation

$$\cos^2 2\theta_e + \cos^2 2\theta_\mu + \cos^2 2\theta_\tau = 1. \qquad (29)$$

At realistic level, there are small deviations of the CL mass-hierarchy angles from benchmark values (27) and (28); they are represented by the shown above data values of CL masses [10]

$$\cos^2 2\theta_e = m_e /(m_e + m_\mu + m_\tau) \cong 2.71 \times 10^{-4}, \qquad (30)$$

$$\cos^2 2\theta_\mu = m_\mu /(m_e + m_\mu + m_\tau) \cong 0.05611, \qquad (31)$$

$$\cos^2 2\theta_\tau = m_\tau /(m_e + m_\mu + m_\tau) \cong 0.94362. \qquad (32)$$

The magnitudes of the CL hierarchy angles follow from (30)–(32),

$$\theta_e \cong 44.53^o, \theta_\mu = 38.15, \theta_\tau = 6.87^o. \qquad (33)$$

To conclude, the results (33) for charged leptons and (14) for neutrinos show that the analogy between CL mass hierarchy angles and neutrino mixing angles is not only by leading approximation bimaximal form (28). It includes the common features of two large and one small angle and close magnitudes of the two large angles.

**2)** Up-quark mass hierarchies

We use PDG values [10] for approximate quark masses

$$m_u \sim 2–3 \text{ MeV}, m_d \sim 4–5 \text{ MeV}, m_s \sim 90 - 100 \text{ MeV}, \qquad (34)$$

$$m_c \sim 1300 \text{ MeV}, m_b \sim 4.5 \text{ GeV}, m_t \sim 174 \text{ GeV}. \qquad (35)$$

Similar to the CL case, the quark benchmark hierarchy-angles are represented by bimaximal patterns of the form (27), (28), while the realistic up-quark hierarchies are

$$\cos^2 2\theta_u = m_u / (m_u + m_c + m_t) \cong 1.4 \times 10^{-5}, \qquad (36)$$

$$\cos^2 2\theta_c = m_c / (m_u + m_c + m_t) \cong 7.4 \times 10^{-3}, \qquad (37)$$

$$\cos^2 2\theta_t = m_t / (m_u + m_c + m_t) \cong 0.9926. \qquad (38)$$

Absolute values of the up-quark hierarchy-angles are given by

$$\theta_u \cong 44.9^o, \ \theta_c = 42.5^o, \theta_t = 2.5^o. \qquad (39)$$

Again, the rule of ‘two large and one small angle and close magnitudes of the two large angles’.

**3)** Down-quark mass hierarchies

Similar to the up-quarks, the realistic down-quark mass hierarchies are given by

$$\cos^2 2\theta_d = m_d / (m_d + m_s + m_b) \cong 9.8 \text{ x } 10^{(-4)}, \quad (40)$$

$$\cos^2 2\theta_s = m_s / (m_d + m_s + m_b) \cong 0.021, \quad (41)$$

$$\cos^2 2\theta_b = m_b / (m_d + m_s + m_b) \cong 0.978, \quad (42)$$

The magnitudes of down-quark hierarchy-angles are

$$\theta_d \cong 44.1^o, \theta_s = 40.8\ ^o, \theta_b = 4.3\ ^o. \quad (43)$$

The main inferences for all SM Dirac mass hierarchy angles is that they are solutions of the Metric-Pythagorean equation of common form – two large close angles plus one small angle similar to the empirical neutrino mixing angles.

**4)** Color quark mass hierarchy

From color symmetry at fixed flavor, the degenerate quark mass hierarchy is represented by one universal hierarchy-angle

$$\cos^2 2\theta^c = m_c / (3m_c) = 1/3, \ \theta^c \cong 27.4^o. \quad (44)$$

That phenomenological hierarchy pattern is an evident universal geometric solution of the Metric-Pythagorean equation of the type (5).

## 7. Zero neutrino mass scale

To date neutrino mass hierarchy is unknown. It is a shortcoming of SM neutrino physics that needs research. In accordance, it is interesting to consider the relations (27) for a connection between neutrino mass hierarchies and neutrino mixing angles. From compatible with world data [7] maximal value of atmospheric neutrino mixing angle $\theta_{23} = 45^o$ they predict zero magnitude of the smallest ‘solar’ neutrino mass $m_1$.

New zero neutrino mass physics with $m_1 = 0$ is especially interesting since it predicts the complete neutrino mass spectrum, neutrino mixing angles and points to Dirac neutrino type.

If the two smallest neutrino masses in the solar mass squared difference are hierarchical $m_2 >> m_1 \cong 0$, neutrino mass ordering is 'normal' and of Dirac type. The complete neutrino mass spectrum is fully determined by the two experimentally known neutrino oscillation mass squared differences,

$$\Delta m^2_{sol} = (m_2^2 - m_1^2) \cong 7.6 \text{ x } 10^{-5} \text{ eV}^2,\ \Delta m^2_{atm} = (m_3^2 - m_2^2) \cong 2.5 \text{ x } 10^{-3} \text{ eV}^2, \quad (45)$$

and is given by[7]

$$m_1 \cong 0,\ m_2 \cong 0.009 \text{ eV},\ m_3 \cong 0.05 \text{ eV}. \quad (46)$$

The main physical characteristics of that unique neutrino mass spectrum (46) are

1) Mass ratios follow

$$m_2 / m_1 >> 1,\ m_3 / m_2 \cong 6. \quad (47)$$

2) So, neutrino mass spectrum (46) is hierarchical and is in complete analogy with the well known from data in the SM hierarchical Dirac spectra of charged leptons and up- and down-quarks, e. g.

$$m(\text{muon}) / m(\text{electron}) \cong 200,\ m(\text{tau}) / m(\text{muon}) \cong 16. \quad (48)$$

3) The sum of neutrino masses is

$$\Sigma m_k \cong 0.06 \text{ eV}. \quad (49)$$

4) Neutrino mixing angles are given by

$$\cos^2 2\theta_{23} = m_1 / \Sigma m_k = 0,\ \cos^2 2\theta_{12} = m_2 / \Sigma m_k = 0.15,\ \cos^2 2\theta_{13} = m_3 / \Sigma m_k = 0.83, \quad (50)$$

$$\theta_{23} \cong 45^o,\ \theta_{12} \cong 33.6^o,\ \theta_{13} \cong 11^o. \quad (51)$$

The magnitudes of soar and atmospheric angles in (51) are in good agreement with known data [7], but the reactor angle disagrees with new data [3-6]. This problem is considered in Sec. 3 with conclusion that the initial Metric-Pythagorean equation (1) may be not exactly applicable for neutrino mixing, but it is successfully applicable at leading benchmark approximation

$$\Delta m^2_{sol} = 0,\ m_2 = m_1 = 0,\ \theta_{23} = 45^o,\ \theta_{12} = 45^o,\ \theta_{13} = 0. \quad (52)$$

Confirmation by new data of the mass sum result (49) in astrophysical and cosmological observations and terrestrial experiments will mean not only discovery of absolute neutrino masses, but also discovery of Dirac neutrino nature.

---

[7] Note that the condition $m_1 = 0$ as a possibility is mentioned in different neutrino phenomenology reviews e. g. [12].

It should be noticed that the empirical messages (46) – (49) suggest a new basic mass scale in elementary particle mass physics – zero neutrino mass scale. Unfortunately, basic theory of zero mass neutrino scale is not developed to date in contrast to empirically not enough supported, but much favorable by theory see-saw mechanism.

If the result (49) for sum of neutrino masses is disproved, the two solar masses must be nearly degenerate $m_2 \cong m_1$ with neutrino mass ordering normal, inverted or quasi-degenerate in contrast to known Dirac particle mass patterns. In that case the neutrinos are likely of Majorana type as the only alternative to Dirac type. Thus the question of Dirac or Majorana neutrino type is related to the factual value of smallest neutrino mass $m_1$.

It should be noticed that there is another possibility of zero neutrino mass in case of 'inverted' mass ordering, $m_3 = 0$, where the neutrino mass spectrum is also completely determined, e. g. at leading approximation

$$m_3 = 0,\ m_1 = m_2 \cong 0.05 \text{ eV},\ \Sigma m_k \cong 0.1 \text{ eV}. \qquad (53)$$

We do not consider this case further because it does not agree with data solar and atmospheric neutrino mixing angles already at the leading approximation.

## 8. Conclusions

Numerous confirmations may indicate that, besides the problem of many free empirical parameters, the SM is a complete theory of elementary particle interaction at low energies not crying for new physics. But there are serious problems. The wonderful, almost universal particle flavor mass and mixing angle hierarchies, that originated from the Higgs sector empirical parameters, are strange in the SM. And there are basic empirical flavor related coincidences that beg for attention.

Firstly, number '3' of SM particle generations coincides with number '3' dimensions of the outer euclidean space. And secondly, the considered in Sec. 3 empirical Pythagorean equation for neutrino mixing angles coincides with the Metric equation for unit vector direction angles in euclidean 3-space geometry.

Those are two basic evidences persistently pointing to connection between SM flavor phenomena and geometry of outer euclidean 3-space. Interestingly, it is supported also by the mathematical fact of two different types of basic solutions without external parameters of the Metric equation (1) – violating geometric symmetry bimaximal solution e.g. (2), and symmetric degenerate solution (5). By experimental data, just these two type solutions are appropriate for neutrino and quark mixing, respectively, at leading approximations.

The SM flavor mass ratios and mixing angles are parts of particle mass matrices and so they are parts of the general relativity gravity source. Probably the noted above low energy nonrelativistic flavor-geometric call for new physics appears a relic from high energy unification of three SM particle generations and three SM basic interactions with geometric gravity of Einstein's general relativity.

The factual SM hierarchy regularities at leading (benchmark) approximation appear nearly universal and simple. At that approximations all known SM Dirac mass hierarchies are united by universal bimaximal form $2\theta k = (90, 90, 0)^{o}$ of the mass hierarchy angles θk for all three Dirac particle flavor mass triads. The singled out '0' in that hierarchy-angle triads correspond to the three largest charged lepton and quark masses, while the two '90' correspond to the two relatively small masses approximated by zeros at that approximations.

The present research reveals noticeable unifying and suggestive power of the Metric-Pythagorean equation in the field of SM Higgs sector parameter regularities. It is essentially supported by Nature's choice of these empirical parameters. The common facts that neutrino mixing angles are nearly bimaximal (e.g. not biminimal) and that all the three types of Dirac particle masses are nearly biminimal (e.g. not bimaximal) is crucial here.

Three color quark mixing angles and one color neutrino mixing angles are united at basic level by 4-color symmetry of the Metric-Pythagorean equation (16), namely all four one-color particle mixing angles are solutions of that primary equation. Different quark and neutrino mixing at the observational level are mainly due to the difference in color numbers of lepton and quark particles.

The breaking of the initial quark-neutrino 4-color symmetry of Eq. (16) does not violate the quark 3-color symmetry[8] at every particular step of the analysis. Particle color matters in the flavor-geometric unification of the very different quark and lepton mixing patterns. Important role of the empirical particle color degree of freedom in the weak interactions appears in the considered phenomenology for the first time.

[8] Empirical quark color in the considered phenomenology is not directly related to the theoretical QCD quark color theory.

In general, it is for the first time that the leading approximations of all known in the SM Higgs sector related empirical mass and mixing hierarchies may be united by basic solutions of one geometric Metric-Pythagorean equation.

Neutrino mixing angle hierarchy is special. In contrast to quark mixing, it appears as spontaneous violation of the initial mixing angle symmetry of the solution-set (2)-(4) of the Metric-Pythagorean equation (1) by singling out one bimaximal solution (2),(7).

Experimentally verifiable indications for neutrino mass spectrum of the same type as charged lepton and quark Dirac spectra are discussed. The magnitude of the smallest neutrino mass value $m_1$ with inequality $m_1 \ll m_2$ should determine the main physical features of that mass spectrum. It points to new zero neutrino mass scale, determines the complete neutrino mass spectrum, $m_1 \cong 0$, $m_2 \cong 0.009$ eV, $m_3 \cong 0.05$ eV, and points to Dirac neutrino type similar to other Dirac mass patterns of charged leptons and quarks.

The other case of nearly degenerate masses $m_1 \cong m_2$ points to Majorana neutrino type with normal or inverted mass hierarchy or quasi-degenerate neutrino masses.

Thus the magnitude of neutrino mass $m_1$ and relation between $m_1$ and $m_2$ are crucial for neutrino mass physics in considered phenomenology. Experimental quantitative confirmation in astrophysical, cosmological and terrestrial measurements of neutrino mass sum $\Sigma m_k \cong 0.06$ eV should be an interesting indication of Dirac neutrino type and absolute neutrino masses. Neutrino mass sum above that special value should point to Majorana neutrino type.